\documentclass[a4paper]{jpconf}

\usepackage{graphicx,color}
\usepackage{amssymb}
\usepackage{amsmath}

\begin{document}
%------------------------------------------------------------------------------ 
% Title
%------------------------------------------------------------------------------ 
%\title{Metal-insulator quantum phase transition 
%in the 1D half-filled Edwards fermion-boson transport model}
%\title{Metal-insulator quantum phase transition 
%in the Edwards transport model}
\title{Dynamic density-density correlations in interacting
Bose gases on optical lattices}
%\title{Metallic phase at the SDW-CDW transition in the
%1D half-filled Holstein-Hubbard model}
%\title{Characterisation of the intervening metallic phase at
%the CDW-SDW transition in the Holstein-Hubbard model}
%------------------------------------------------------------------------------ 
% Authors
%------------------------------------------------------------------------------ 
\author{S.\ Ejima$^1$, H.\ Fehske$^1$, and F.\ Gebhard$^2$}
\address{
Institut f{\"ur} Physik,
Ernst-Moritz-Arndt-Universit{\"a}t Greifswald,
17489 Greifswald,
Germany}
\address{
Fachbereich Physik,
Philipps-Universit{\"a}t Marburg,
35032 Marburg,
Germany
}
%------------------------------------------------------------------------------ 
% E-mail
%------------------------------------------------------------------------------ 
\ead{<ejima,fehske>@physik.uni-greifswald.de}

%------------------------------------------------------------------------------ 
% Date
%------------------------------------------------------------------------------ 
\date{\today}

\begin{abstract}
In order to identify possible experimental signatures of the superfluid to Mott-insulator quantum phase transition
we calculate the charge structure factor $S(k,\omega)$ for the one-dimensional Bose-Hubbard model using the dynamical density-matrix renormalisation group (DDMRG) technique.
Particularly we analyse the behaviour of $S(k, \omega)$
by varying---at zero temperature---the Coulomb interaction strength within the first Mott lobe.
For strong interactions, in the Mott-insulator phase, we demonstrate that the DDMRG  
results are well reproduced by a strong-coupling expansion, just as the quasi-particle
dispersion. In the super\-fluid phase we determine the linear excitation spectrum
near $k=0$. 
%compare the DDMRG data with results from mean-field theory.
In one dimension, the amplitude mode is absent
which mean-field theory suggests for higher dimensions.
\end{abstract}

Recent experimental realisations of optical lattices make it
possible to investigate the properties of ultracold dilute atoms
in a new regime of strong
correlations~\cite{GMEHB02,BDZ08}. 
Tuning the strength of the laser field, the effective interactions 
between the atoms can be 
%achieved
tuned
to be stronger than their kinetic energy. The competition between
the Coulomb energy and the kinetic energy may even drive a quantum phase
transition between superfluid (SF) and Mott insulating (MI) phases.
The Bose--Hubbard model (BHM) captures the essential physics 
of this problem. In previous work~\cite{EFG11} we studied 
the photoemission spectra of the one-dimensional (1D) BHM both 
in the MI and SF phases. In this contribution we extend our preceding  
investigations of the dynamical properties of the BHM by 
analysing the dynamical structure factor $S(k,\omega)$. 
Analytically, $S(k,\omega)$ has been determined 
%simple 
% drop this word
by mean-field approaches~\cite{ODFSS05,HABB07} and, numerically, 
for rather small 1D systems, by exact diagonalisation~\cite{RB04} or, 
at finite temperatures, by QMC~\cite{PEH09}.  Here, 
we carry out large-scale dynamical density-matrix renormalisation 
group (DDMRG)~\cite{DDMRG} calculations in order to examine the density-density 
correlations of  $N$ interacting Bose particles on $L$ sites  
at zero temperature. Moreover, fixing $\rho=N/L=1$, we set 
up a perturbation theory whose results can be 
compared with the DDMRG data in the strong-coupling limiting case.

The Hamiltonian of the 1D BHM reads 
\begin{eqnarray}
\hat{{\cal H}}=
 -t\sum_j( \hat{b}_j^\dagger \hat{b}_{j+1}^{\phantom{\dagger}}
 +\hat{b}_{j}^{\phantom{\dagger}}\hat{b}_{j+1}^\dagger )
 +\frac{U}{2}\sum_j \hat{n}_{j}(\hat{n}_{j}-1)\,,
\label{bhm}
\end{eqnarray}
where $\hat{b}_j^\dagger$ and $\hat{b}_j^{}$ are the creation and
annihilation operators for bosons on site $j$, 
$\hat{n}_j=\hat{b}_j^\dagger\hat{b}_j^{}$ 
is the boson number operator on site $j$. 
The tunnel amplitude between nearest neighbour lattice
sites is denoted by $t$, and $U$ gives the on-site Coulomb repulsion of the 
Bose particles. For the DDMRG treatment of~(\ref{bhm}) 
we use periodic boundary conditions, take into account $n_b=5$ bosons  
per site and keep up to $m=500$ density-matrix eigenstates, ensuring  
that the discarded weight is always smaller than $3\times10^{-5}$.

% S(k,\omega)
The dynamical charge structure factor is defined as
\begin{eqnarray}
 S(k,\omega)=\sum_n|\langle\psi_n|\hat{n}_k|\psi_0\rangle|^2
             \delta(\omega-\omega_n)\,,
\end{eqnarray}
where $|\psi_0\rangle$ and $|\psi_n\rangle$ denote the ground and 
$n$-th excited state, respectively, and 
$\omega_n=E_n-E_0$ gives the corresponding excitation energy. 
$S(k,\omega)$ characterises the density-density response
of the Bose gas and is directly accessible in experiment, for instance, by Bragg
spectroscopy~\cite{SICSPK99}. 

\begin{figure}[b]
 \begin{center}
  \includegraphics[width=.9\linewidth]{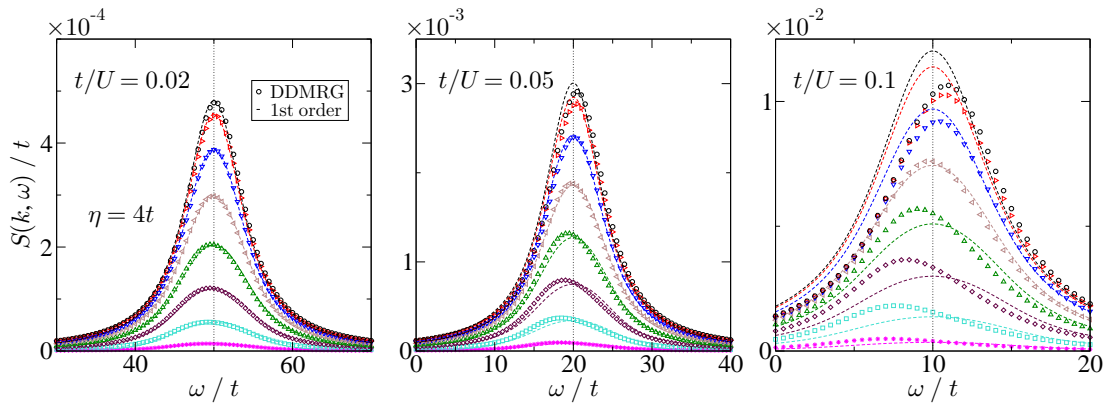}
 \end{center}
 \caption{
 Frequency dependence of the dynamical charge structure factor $S(k,\omega)$
at various momenta $k=2\pi j/L$ (for $j=1,2,\ldots,L/2$ from bottom to top).
 DDMRG data (symbols) were obtained for $L=16$; $S^{(1)}_{\eta}(k,\omega)$ (lines) 
give the corresponding first-order strong-coupling expansion results 
 (for the same value of $\eta=4t$). The dashed line indicates the 
 point $\omega=U$.
 }  
 \label{fig1}
\end{figure} 

From %first-order 
lowest-order % the result is the simplest=leading-order result which one 
%can get; it is not so easy to be systematic in the order
strong-coupling perturbation theory we find
for $\rho=N/L=1$ ($N,L\to\infty$)
\begin{eqnarray}
S^{(1)}(k,\omega>0)=\left(\frac{4t\sin(k/2)}{U}\right)^2 
\int_{-\pi}^{\pi}\frac{{\rm d}k}{\pi}\sin^2k \,
\delta(\omega-U+2t\cos k\sqrt{5+4\cos k})\,.
\end{eqnarray}
Then, for $U-2\sqrt{5+4\cos k}\leq\omega\leq U+2\sqrt{5+4\cos k}$,
we have in units of $t$
\begin{eqnarray}
 S^{(1)}(k,\omega>0)
 =\left[\frac{4\sin(k/2)}{U}\right]^2
  \frac{\sqrt{20+16\cos k-(\omega-U)^2}}{2\pi(5+4\cos k)}\,,
 \label{sc_unconvolved}
\end{eqnarray}
where $0\leq k<2\pi$. 
Since DDMRG provides $S(k,\omega)$ with a finite broadening $\eta$,
it is useful to convolve the strong-coupling result $S^{(1)}(k,\omega)$
with the Lorentz distribution~\cite{NE04}:   
% convolution
%\begin{eqnarray}
$ S_{\eta}(k,\omega)
 =\int_{-\infty}^{\infty}{\rm d}\omega^{\prime} S(k,\omega^{\prime})
   \eta/\left[\pi\left[(\omega-\omega^\prime)^2+\eta^2\right]\right]$.
%\end{eqnarray}
For the 1D BHM, from Eq.~(\ref{sc_unconvolved}), we thus get 
\begin{eqnarray}
 S_{\eta}^{(1)}(k,\omega>0)= \left(\frac{4\sin(k/2)}{U}\right)^2 
  \frac{2}{\pi^2}\int_{-1}^{1}{\rm d}\lambda 
  \frac{\eta\sqrt{1-\lambda^2}}
       {(\omega-U+2\lambda\sqrt{5+4\cos k})^2+\eta^2}\,.
\end{eqnarray}

Figure~\ref{fig1} illustrates the frequency and momentum dependencies
of $S(k,\omega)$. Peaks in $S(k,\omega)$ assign charge excitations. 
For $k\simeq 0$, $S(k,\omega)$ exhibits a maximum around $\omega\approx U$
that can be attributed to excitations across the Mott gap. 
Within strong-coupling theory~(\ref{sc_unconvolved}) this signature 
stays at $\omega\approx U$ for all $k$ but becomes more pronounced 
as $k$ reaches the Brillouin zone boundary. Our DDMRG data corroborate
this prediction (cf.\ the left panel of Fig.~\ref{fig1} depicting 
the results for $t/U=0.02$). Naturally, as $U$ gets smaller, differences
between the numerical DDMRG and the analytical strong-coupling
results emerge. Most importantly, the position of the maximum in  
$S(k,\omega)$ varies with $k$: it is shifted to smaller (larger) frequencies
for $k$ values near the band centre (Brillouin zone boundary); 
see middle and right-hand panels of  Fig.~\ref{fig1}. 
We have to include higher-order $t/U$--corrections 
to reproduce this feature analytically. 
Note that the maximum in $S(k,\omega)$ amplifies as $k\to \pi$. 

\begin{figure}[t]
 \begin{center}
  \includegraphics[width=.92\linewidth]{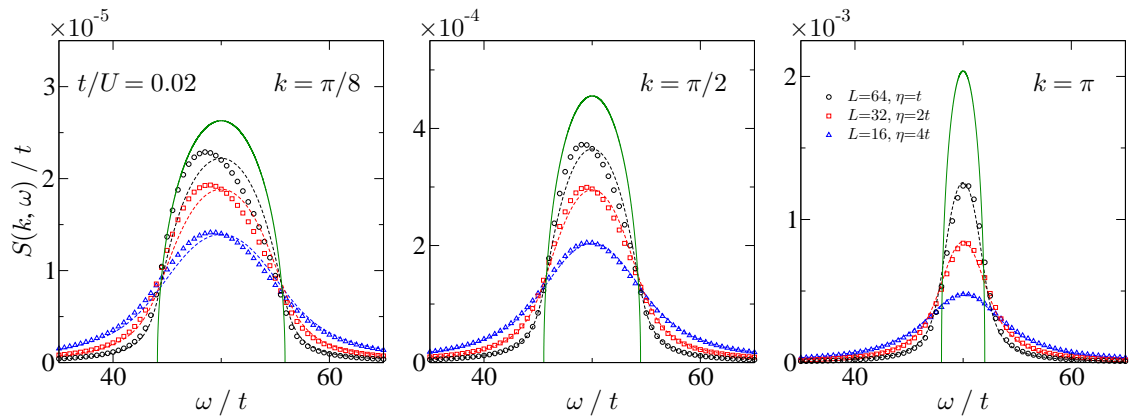}
 \end{center}
 \caption{
 $\eta$-dependence of the dynamical structure factor $S(k,\omega)$. 
 DDMRG data (symbols) are compared with the corresponding 
 strong-coupling results (dashed lines) at various 
momenta~$k$. The solid green line shows the unconvolved result of the 
 strong-coupling expansion.
 }  
 \label{fig2}
\end{figure}

Next, we investigate the dependences of $S(k,\omega)$ on the
broadening $\eta$ and the system size $L$ 
to scrutinise whether the DDMRG data ``converge'' to the 
unconvolved strong-coupling result~(\ref{sc_unconvolved})
as $\eta\to 0$ and $L\to\infty$. Figure~\ref{fig2}---showing $S(k,\omega)$ 
in the MI phase at $k=\pi/8$ (left), $\pi/2$ (middle) and $\pi$ (right panel)
for $L=16$, 32, 64 $\eta=4t$, $2t$, $t$, respectively---demonstrates that 
this is indeed the case. Firstly, for $t/U=0.02$, we see that
the DDMRG results are in a satisfactory accord with 
$S_{\eta}^{(1)}(k,\omega)$, where 
as a matter of course for smaller system sizes a larger value 
of $\eta$ has to be used to achieve good agreement. Secondly,
$S(k,\omega)$ calculated by DDMRG approaches the curve 
given by~(\ref{sc_unconvolved}) for increasing system size $L$ and 
decreasing broadening $\eta$.

 \begin{figure}[t]
 \begin{center}
  \includegraphics[width=.79\linewidth]{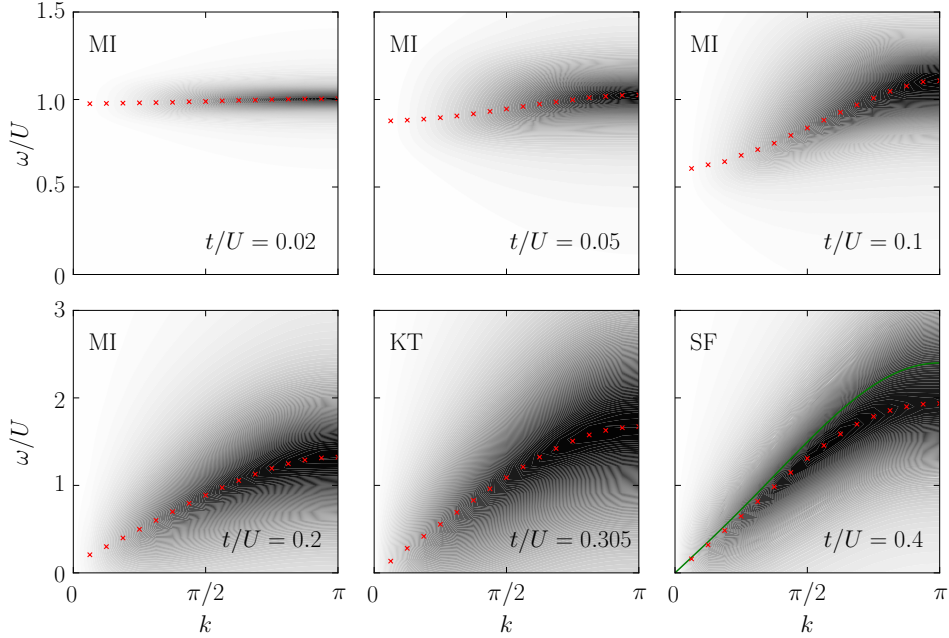}
 \end{center}
 \caption{
 Intensity of the dynamical structure factor $S(k,\omega)$ 
 in the $k$--$\omega$ plane for different 
 $t/U$. DDMRG data were obtained for $L=32$ system, using  
% $\eta=2t$ and $t$ in the MI and SF phases, respectively.
 $\eta=2t$ ($\eta=t$) in the MI phase and at the KT transition point
 (in the SF phase).
 The red crosses mark the positions of the maximum in each
 $k=2\pi j/L$--sector.
 The green line for $t/U=0.4$ marks the field-theory/Bogoliubov result 
 $\omega(k)=\sqrt{\epsilon_0(k)[\epsilon_0(k)+2U]}$ with
 $\epsilon_0(k)=-2t(\cos k-1)$.
  }
 \label{fig3}
\end{figure} 

Finally, we look at the changes in the dynamical density-density 
response as the system crosses the MI-SF quantum phase transition
with decreasing Coulomb interaction strength. Figure~\ref{fig3}
shows the intensity distribution of $S(k,\omega)$ in 
the MI and SF phases as well as in the vicinity of 
the Kosterlitz--Thouless (KT) 
transition point, where the charge excitation gap closes.
For large $U$, the spectral weight is mainly concentrated around $\omega=U$
in the region $\pi/2\leq k\leq \pi$, in  
agreement with the strong-coupling prediction.
%If 
As % it is not a condition, just `as it happens that'
$U$ weakens in the MI phase, the distribution of the 
spectral weight broadens. At the same time, the maximum 
value of $S(k,\omega)$ acquires a sizable dispersion 
(see upper-row panels). As the system reaches the MI-SF
transition point we observe a significant redistribution
of spectral weight to lower $k$ values and, most notably, the 
excitation gap closes (see lower panels). In our previous work,
we evaluated the scaling of the Tomonaga--Luttinger liquid
parameter and determined the KT transition point
to be located at $t/U=0.305(1)$~\cite{EFG11}. 
In the SF phase 
spectra for $k\gtrsim 0$ (cf.\ the panel for $t/U=0.4$), we observe an almost 
linear dispersion of the $S(k,\omega)$, which---in accordance 
with bosonization~\cite{CCGOR11} and 
with Bogoliubov theory~\cite{Bo47} 
---can be taken as a signature of 
the Bose condensation process. 
%Aside let us point out
%We note in passing that % Florian prefers this phrase
Our 1D DDMRG BHM data are unsuggestive of two distinct
(gapless sound and massive) modes in the SF phase. 
%as found by the mean-field treatment of the 2D BHM system~\cite{HABB07}.
These two phases are found in mean-field theory and may appear 
for dimensions $d\geq 2$ where a true condensate exists
in the SF ground state.

To summarise, we have determined the dynamical structure factor 
$S(k,\omega)$ for the 1D BHM with particle density 
% one
$\rho=N/L=1$ % that's clearer
by means of unbiased numerical DDMRG calculations. 
% Likewise the previously discussed photoemission spectra~\cite{EFG11}, 
As discussed for photoemission spectra previously~\cite{EFG11}, 
% sounds better this way
$S(k,\omega)$ agrees with the first-order perturbation theory result
in the Mott insulator phase for $U\gg t$. Naturally, as the Coulomb 
interaction is lowered, noticeable deviations appear between both approaches, 
in particular the DDMRG $S(k,\omega)$ becomes dispersive and 
we find a substantial redistribution of spectral weight into 
the small $k$-sector. %Obviously, %can be taken out to save space
In this regime, 
higher-order corrections have to be taken into account 
in our analytical treatment of the BHM.   
Approaching the SF state, the charge excitations gap closes and
the maximum in $S(k,\omega)$ exhibits a linear dispersion. 
The quantum phase transition between MI and SF phases is 
located at about $t/U\simeq 0.3$ and found to be of
KT type.

{\it Acknowledgements.} 
SE and HF acknowledge funding by DFG through SFB 652.

%%% Bibliography %%%
\section*{References}
%\bibliography{ref}
%\bibliographystyle{jphysicsB} 
%\bibliographystyle{unsrt} 
%\bibliographystyle{plain}
%\bibliographystyle{apsrev}
\bibliographystyle{iopart-num}
\providecommand{\newblock}{}

\end{document}